\documentclass[10pt]{article}
\usepackage{latexsym}
\usepackage{epsfig}
\begin{document}

\title{Spectra and M1 Decay Widths of Heavy-Light Mesons}

\author{T.A. L\"ahde\footnote{talahde@rock.helsinki.fi},$\:$
 C.J. Nyf\"alt\footnote{nyfalt@rock.helsinki.fi}$\:$ and
D.O. Riska\footnote{riska@science.helsinki.fi}}

\date{February 22, 2000}

\maketitle

\centerline{\it Department of Physics, POB 9, 00014 University of
Helsinki, Finland}

\vspace{1cm}

\begin{abstract}
The M1 decay widths of charm and bottom 
mesons and their excited states
are sensitive to the
relativistic aspect of the quark current operators
and obtain a significant contribution from the interaction current
that is associated with the scalar
confining interaction between heavy quarks ($Q$) and light antiquarks
($\bar q$).
Consequently they
provide direct evidence on the latter.
The spectra and the M1 transition widths of the $D$, $D_s$ and $B$,
$B_s$ mesons and
their orbital excitations are calculated here within the framework of 
the Blankenbecler-Sugar equation,
which allows a covariant treatment, while retaining a formal
similarity to the nonrelativistic approach. 
The hyperfine interaction
in the $Q\bar q$ mesons is modeled by screened
gluon exchange, which shares many features with the 
instanton induced interaction.
The quality of the model is tested by calculation of the spectra
and M1 decay widths of charmonium.
  
\end{abstract}

\newpage

\section{Introduction}
\vspace{0.5cm}

A dynamical description of the spectrum of the heavy flavor
mesons $Q\bar q$ has {\it a priori} to be relativistic as the 
confined light quarks and antiquarks have velocities close to that 
of light. Combination of this
requirement, with the nonrelativistic instantaneous aspect of the
effective linear \cite{Bali} (or near linear \cite{Diak}) 
confining interaction 
suggests that a 3-dimensional
quasipotential reduction of the Bethe-Salpeter equation should provide
the most obvious framework for a realistic description. This is 
{\it a fortiori}
the case as the Bethe-Salpeter equation with instantaneous
interactions, while commonly employed, \cite{Metsch}, fails to provide
satisfactory results for the M1 transitions of the analogous 
heavy quarkonia \cite{Snellman}.

Among the many choices of quasipotential equations the
Blankenbecler-Sugar equation \cite{Blankenbecler,Logunov} suggests 
itself because of its transparent formal similarity to the
nonrelativistic Lippmann-Schwinger
equation, to which it reduces in the adiabatic limit.
Moreover it admits direct employment of the conventional
(relativistic) models for the hyperfine interaction between quarks 
and antiquarks, as well as the linear confining interaction   along
with its obvious relativistic modifications. The key advantage
of the Blankenbecler-Sugar equation over the nonrelativistic
or ``relativized'' versions of the nonrelativistic
wave equation is the unapproximated treatment of the
relativistic two-particle kinetic energy operator of the $Q\bar q$
system. The relativistic modifications to the interaction are
in comparison modest, and may be treated in an approximate
fashion, when convenient. This feature is shared with the
Gross \cite{Gross} reduction of the Bethe-Salpeter
equation, which has been applied to heavy flavor mesons
in ref.\cite{Wally}.
 
Given this situation, several different versions of the 
Blankenbecler-Sugar
equation have been employed in the literature \cite{Coe}. We shall
here use the symmetrical version of ref. \cite{Lomon}
to calculate both the spectra and the M1 decay widths of
the $D$ and $D_s$  as well as the $B$ and $B_s$ mesons and their 
orbital excitations.
The formalism
then differs somewhat from that used to describe the $Q\bar q$ system
in refs. \cite{Ebert}. 
The main advantage of the version
used here is that it leads to a Lippmann-Schwinger type wave-equation.
It differs most obviously from that used in refs.\cite{Ebert}
in the appearance of the ``minimal relativity'' \cite{Jackson} square 
root factors $\sqrt{m/E}$ in the effective interaction
potential in the wave equation.

The radiative transitions of the ground state $Q\bar q$ mesons
may be described phenomenologically by effective field theory
methods \cite{Amund,Ang,Bajc}. A complete dynamical model is
however required for any comprehensive description of the decays
of the full spectrum of $Q\bar q$ states. The static quark
model is insufficient here, as it is expected to overpredict the 
width of the M1 transition for 
$D^{+*}\rightarrow D^+\gamma$ by
a large factor in analogy to its 
large overprediction of the width for the M1 decay 
$J/\psi\rightarrow \eta_c\gamma$, and which is
resolved only by employment of the relativistic magnetic
moment operator for the heavy quarks along with the
interaction current operator that is associated with the
scalar confining interaction \cite{Timo}.

In comparison to heavy quarkonia the heavy-light $Q\bar q$
mesons are more complex systems in view of the still unsettled
form of the hyperfine interaction for the light quarks
and concomitantly their empirically still only poorly
known spectrum. While the branching ratios of the M1 decays
of the charmed vector mesons are known, their still
unknown total decay widths prevent an absolute empirical
determination of their M1 decay widths.
In spite of -- and  because of --
this situation it remains a challenge to decode the
structure and the dynamics of these systems, as this is a
main key to the understanding of the nature of the confining
interaction and the hyperfine interaction between 
constituent quarks. 

There are good theoretical \cite{Bali,Gromes} and phenomenological
\cite{Timo} reasons to expect the confining interaction to be
linear with a scalar spinor structure, although mixtures of
combinations of scalar and vector coupling terms continue
to be employed \cite{Metsch,Ebert}. For the heavy light
$Q\bar q$ systems the hyperfine interaction is likely to
be formed as a combination of the screened
perturbative gluon exchange
interaction
\cite{Godfrey} and the direct instanton induced interaction
\cite{Nowak}, which share many qualitative features. 
The latter appears in the instanton liquid
model description of the QCD vacuum \cite{Rho}, which is
supported by numerical lattice calculations 
\cite{Negele1,Negele2}. In the case of heavy light quarkonia
the relativistic corrections to the gluon exchange
interaction cause an appreciable damping of that
interaction, besides the screening of the the color
hyperfine coupling strength.

We here consider the M1 decay rates of the $Q\bar q$ systems,
because these are expected to be overpredicted by the static 
quark model, and are likely to provide decisive 
information on the nature of the confining interaction.
Because the constituent quark masses appear in
different combinations in the magnetic transition
operators for the different charge states of the
$Q\bar q$ mesons the M1 transition rates should also
provide information on the constituent quark masses.
We show that the decay rates predicted by  
the static quark model for the
M1 transition $D^*\rightarrow D\gamma$ 
is strongly reduced with (1)
employment of the relativistic magnetic moment operator
for the quarks in combination with 
(2) the interaction current,
that appears if the confining interaction couples as 
a \mbox{scalar}.
This result falls in line with that for the
corresponding M1 decay rate problem of heavy quarkonia
\cite{Timo}. For $q\bar q$ systems (i.e. systems with
equal quark and anti-quark masses) neither the
gluon exchange interaction nor the instanton
induced interaction gives rise two any two-body
current operator with a spin-flip part. The spin-flip
component of those exchange current operators that
appear in the case of $Q\bar q$ systems usually have much
smaller matrix elements compared to the matrix
elements of the two-body current that is associated with
the scalar confining interaction.

The effective interaction potential between the heavy
quark and the light antiquark is accordingly described
by a scalar linear confining 
+ one-gluon exchange
interaction with a hyperfine interaction
formed of the perturbative
gluon exchange interaction. The parameters of this model
are, to the extent possible, fixed to values obtained by
lattice calculations as well as to the known part of the
$D$ and $D_s$ meson spectra, and then appropriately scaled
to the bottom mesons. Finally the quality of the model
is tested by a calculation of the spectrum and the M1
decay widths of charmonium. 

With the framework of the Blankenbecler-Sugar equation
and the $Q\bar q$ interaction models above it
is possible to obtain spectra for both charmonium and
the heavy-light mesons, which agree overall with the empirical
spectra, where these are known. Similarly the description
of the (known) M1 decay rates of the $J/\psi$ and 
the $\psi\,'$ are well described. The M1 decay rates for
the $D$ and $B$ mesons are predictions, as absolute
empirical determination of these are yet to be made.  
The present result that it is possible to achieve a
satisfactory description of both the spectra and M1
decay rates of heavy quarkonia and heavy light mesons
differs from that of ref.\cite{Ebert2}. The key point
here is the employment of an unapproximated quark
current operator.

This report falls into 7 sections. Section~\ref{sec_bslt} 
contains a description of the 
application of the Blankenbecler--Sugar equation 
to the $Q\bar q$ and $Q\bar Q$ systems.
Section~\ref{int_ham} contains a description of the interaction parts
of the Hamiltonian and the current operators.
The calculated $D, D_s$ and $B, B_s$ meson 
as well as the $c\bar c$ spectra
are described in section~\ref{sec_spec}.
Section~\ref{sec_curr} describes the quark current
operators, including the two-quark current that
arises along with a linear scalar confining
interaction. The calculated M1 transition
rates are reported in section ~\ref{sec_M1}. 
Section~\ref{sec_disc} contains a concluding discussion.

\vspace{1cm}

\section{The Blankenbecler--Sugar equation applied to 
$Q\bar q$ systems} \label{sec_bslt}

\vspace{0.5cm}

The Blankenbecler--Sugar equation for a $Q\bar q$ system may be 
expressed 
as an eigenvalue equation of the form
\begin{equation}
\left(\frac{\vec p\,^2 }{2m_r} + V\right)\psi=\epsilon \psi.
\label{wave}
\end{equation}

Here $m_r$ is the reduced mass of the heavy quark ($Q$) mass
$M$ and the light  
the antiquark ($q$) mass $m$: $(m_r=Mm/(M+m))$. The 
the eigenvalue $\epsilon$ is related to the energy $E$ of the
$Q\bar q$ meson system as
\begin{equation}
\epsilon=\frac{[E^2-(M-m)^2][E^2-(M+m)^2]}{8 m_r
 E^2}.
\end{equation}
The interaction operator $V$, which need not be a local
operator, may be obtained from the $Q\bar q$ irreducible
quasipotential ${\cal V}$ (in momentum space) as \cite{Risk}:

\begin{equation}
V(\vec p\,',\vec p\,)=\sqrt{\frac{M+m}{W(\vec p\,')}}
{\cal V}(\vec p\,',\vec p\,)
\sqrt{\frac{M+m}{ W(\vec p\,)}}.
\label{qpot}
\end{equation}

Here the function $W$ is defined as
\begin{equation}
W(\vec p)=E_Q(\vec p\,)+E_{\bar q} (\vec p\,),
\end{equation}
with $E_Q(\vec p\,)=\sqrt{M^2+\vec p\,^2}$ and
$E_{\bar q}(\vec p\,)=\sqrt{m^2+\vec p\,^2}$ respectively.
In the Born approximation the quasipotential ${\cal V}$ equals
the $Q\bar q$ invariant scattering amplitude ${\cal T}$, and
thus a constructive relation to field theory obtains.

Numerical solution of the wave equation~(\ref{wave}) yields a
value for the eigenvalue $\epsilon$. The energy (rest mass)
of the $Q\bar q$ meson is then obtained as
\begin{equation}
E=\sqrt{M^2+m^2+4m_r\epsilon
+2\sqrt{M^2m^2+2m_r(M^2+m^2)\epsilon+4m_r^2\epsilon^2}}.
\end{equation}
In the equal mass case $M=m$ this expression reduces
to $E=\sqrt{4m(m+\epsilon)}$.

The relation between the (total) energy $E$ of the $Q\bar q$
system and the eigenvalue $\epsilon$ in the equation (1)
shows that the relativistic treatment of the kinetic
energy term in the Hamiltonian leads to a lowering of the
former in comparison to what a nonrelativistic treatment
would have implied. This is a consequence of the effective
weakening of the repulsive kinetic energy operator, that
results when the quadratic $p^2$ terms are 
replaced by square
roots of $p^2$. Despite the formal similarity between the
Blankenbecler-Sugar equation and the Schr\"odinger
equation, the former employs a quadratic mass operator.

The main component of the interaction between the $Q$ and 
$\bar q$ quarks is the linear confining interaction, which 
is well defined only in the position space representation, 
which we
shall therefore adhere to. While the Blankenbecler--Sugar
equation takes full account of the relativistic two-quark 
kinetic energy operator, the 
relativistic aspects of the hyperfine interaction operators
will be treated
perturbatively. This treatment is more than adequate for the
heavy quark, but less so for the light one. It does,
however, 
allow retention of the conventional operator structure
of the interaction operator between the quarks.

\vspace{1cm}

\section{The Interaction Hamiltonian for
$Q\bar q$ systems} \label{int_ham}

\vspace{0.5cm}

\subsection{The confining interaction }

The linear confining interaction will here -- on the basis of
compelling indications -- be assumed to be a scalar in the
spinor representation \cite{Timo,Gromes}. To second order
in  the inverse quark masses this interaction, as calculated from the
invariant amplitude according to~(\ref{qpot}), 
takes
the form
\begin{eqnarray}
V_C & = & cr\left(1-\frac{3}{2}\frac{\vec p\,^2}{m_2^2}\right)
+\frac{c}{4 M m r}
-\frac{c}{r}\frac{M^2+m^2}{4  M^2 m^2}\vec S\cdot \vec L \nonumber \\
 & & \mbox{} +\frac{c}{r}\frac{M^2-m^2}{8M^2m^2}
(\vec\sigma_Q-\vec \sigma_{\bar q})\cdot \vec L.
\label{conf}
\end{eqnarray}

Here $c$ is the string tension parameter, the value of
which is $\sim$ 1 GeV/fm.
The mass coefficient $m_2$ is defined as
\begin{equation}
\frac{1}{m_2}=\frac{\sqrt{M^2+m^2+Mm}}{\sqrt{3}Mm}. \label{mass}
\end{equation}
The spin--independent term that is proportional
to $1/r$ is a consequence of the square root factors
in~(\ref{qpot}). Without those factors the factor 3/2
in the momentum dependent term in the first bracket on the
r.h.s, of~(\ref{conf}) would be 1.
The terms of second order in the inverse quark 
masses in~(\ref{conf}) are implied by 
scalar coupling for the confining interaction.
As the antisymmetric spin-orbit interaction has no
diagonal matrix elements between any of the states
in the $S$- and $P$-shells, and only connects spin
singlet and triplet states with $J=L$ it contributes to
the energy only at quartic order in the inverse quark
masses.

The term of second order in $\vec p$ in the spin-independent
term in~(\ref{conf}) is unrealistically large for light quarks, but 
is substantially moderated by inclusion of the next term
in the (asymptotic) $(v/c)^2$ expansion. 
Similarly the spin-orbit interaction in~(\ref{conf}) is 
unrealistically large for light quarks.
Both of these explicitly velocity dependent terms are strongly
moderated by the higher order terms in the (asymptotic)
expansion in $\vec v\,^2$.
This is seen directly by comparison to the corresponding terms
in the 
the quartic correction term to~(\ref{conf}),
which takes the
form

\begin{eqnarray}
V_C^{(4)} & = & -\frac{c\pi}{16m_{4a}^4}\delta^{(3)}(r)
-\frac{c}{r}\frac{\vec p\,^2}{16m_{4b}^4}
-\frac{c}{r}
\frac{(\vec p\,^2-(\hat r\cdot\vec p)^2)}{32 m_{4c}^4} \nonumber \\
& & \mbox{} +cr\frac{3}{8}\frac{\vec p\,^4}{m_{4d}\,^4}
+\left\{\frac{c}{32 r^3}\left[\frac{1}{M^3}\left(\frac{3}{M}+\frac{2}
{m}\right)+\frac{1}{m^3}\left(\frac{3}{m}+\frac{2}{M}\right)\right]
 \right.\nonumber \\
& & \mbox{} \left. +\frac{c}{8 r}\frac{\vec p\,^2}{M^2 m^2}
\left(5+\frac{3}{2}\left(\frac{M}{m}\right)^2
+\frac{3}{2}\left(\frac{m}{M}\right)^2 \right)\right\}\vec S\cdot
\vec L \nonumber \\
&&
+{1\over 16 M^2 m^2}{c\over r^3}Q_{12}.
\label{c4}
\end{eqnarray}
where $Q_{12}$ is the quadratic spin-orbit
interaction operator 
\begin{equation}
Q_{12}= (\vec\sigma_Q\cdot \vec L\,\vec\sigma_{\bar q}\cdot \vec L+
\vec\sigma_{\bar q}\cdot \vec L\,\vec\sigma_Q\cdot \vec L)/2.
\end{equation}
The quartic component of the antisymmetric spin-orbit
interaction has been dropped here, as it contributes to the
energies of the meson states only to sixth order.
Above the mass coefficients $m_{4j}$ are defined as

\begin{eqnarray}
\frac{1}{m_{4a}\,^4}&=&\left[\frac{(M+m)^{2}}{M^{3}m^{3}}
-\frac{1}{M^{2}m^{2}}\right] \\
\frac{1}{m_{4b}\,^4}&=&\left[\frac{3(M+m)^{4}}{M^{4}m^{4}}
-\frac{8(M+m)^{2}}{M^{3}m^{3}}\right] \\
\frac{1}{m_{4c}\,^4}&=&\left[\frac{5(M+m)^{4}}{M^{4}m^{4}}
-\frac{16(M+m)^{2}}{M^{3}m^{3}} + \frac{2}{M^{2}m^{2}}\right] \\
\frac{1}{m_{4d}\,^4}&=&\left[\frac{(M+m)^{4}}{M^{4}m^{4}}
-\frac{3(M+m)^{2}}{M^{3}m^{3}}+\frac{1}{M^{2}m^{2}}\right]
\end{eqnarray}

The scalar structure of the confining interaction implies functional
relations between the different spin components of the potential. Thus
e.g. in the static limit the spin-orbit component may be calculated from
the central component as
\begin{equation}
v_{LS}(r)=-\frac{1}{4}\left(\frac{1}{M^2}+\frac{1}{m^2}\right)\frac{1}
{r}\frac{\partial}{\partial r}v_c(r).
\end{equation}
Here $v_c(r)$ is the central confining
potential and $v_{LS}(r)$ is
the coefficient function for the spin-orbit coupling operator $\vec
S\cdot \vec L$. When the static limit is not invoked, the corresponding
relation is more complicated. To a good approximation it may however be
cast in the form
\begin{eqnarray}
v_{LS}(r) & = & -\frac{2c}{\pi r}
{\partial\over \partial r}
\int_{0}^{\infty}dr'r'^3  \nonumber \\
& & \int_{0}^{\infty} dkk^2j_0(kr)j_0(kr')\left\{\frac{1}{(e_Q+M)^2}+\frac{1}
{(e_{\bar q}+m)^2}\right\},
\end{eqnarray}
where $e_Q=\sqrt{M^2+k^2/4}$ and $e_{\bar q}=\sqrt{m^2+k^2/4}$. This
relation implies that the static approximation to the spin-orbit
interaction in~(\ref{conf}) represents a considerable overestimate of the net
spin-orbit interaction in the case of light constituent quarks.

A similar more accurate representation of the quadratic spin-orbit
component of the confining interaction obtains with the replacement
\begin{eqnarray}
\frac{1}{16 M^2m^2}\frac{c}{r^3}Q_{12} & \rightarrow & -\frac{2c}{\pi
r}{\partial\over \partial r}{1\over r}{\partial\over \partial r}
\int_{0}^{\infty}dr'r'^3 \nonumber \\
& & \mbox{} \int_{0}^{\infty}dkk^2\frac{j_0(kr)j_0(kr')}
{(e_Q+M)^2(e_{\bar q}+m)^2}.
\end{eqnarray}

The $\vec p\,^2$ expansion of the interaction is an
asymptotic series, which has to be truncated. As in 
the case of constituent quarks the successive terms have
increasing amplitudes of alternating signs, this
series is but a poor representation of the full
spinorial structure of the interaction, the velocity
dependence of which is well behaved. We shall therefore
truncate the interaction to second order and retain
only the local spin-orbit and quadratic spin-orbit
interaction terms of quartic order. The latter is
numerically insignificant for the low angular momentum
states considered here.

To this order the most realistic treatment of the
velocity dependent
correction 
$-3/2\vec p\,^2/m_2^2$
to the linear potential $cr$ in~(\ref{conf}) 
would be to take it into account as a
mass shift,
\begin{equation}
m^*\rightarrow m^* + {3\over 2} cr ({m^*\over m_2})^2, 
\end{equation}
in the kinetic term in~(\ref{conf}). 
We shall treat this reduction of the kinetic term
phenomenologically by subtraction of a constant ($b$)
from the kinetic term. The value of this constant is
determined by a fit of the calculated spectra to
the experimental one. Without such a constant the
kinetic term in the Hamiltonian becomes unrealistically 
large as becomes evident below.


\subsection{The one-gluon exchange interaction}
\label{oge}
\vspace{0.5cm}

In the case of heavy quarkonia the perturbative gluon exchange
interaction forms an important component of the hyperfine interaction
between quarks \cite{Godfrey}. To second order in the
inverse quark masses (with exception for the
quadratic spin-orbit interaction, which is quartic) that interaction,
after multiplication of the square root factors in~(\ref{qpot}), takes the
form:
\begin{eqnarray}
V_G & = & -\frac{4}{3}\alpha_s\left\{\frac{1}{r}-\frac{3\pi}
{2m_2^2}\delta^{(3)}(r)+\frac{\vec p\,^2}{ 2 M m r}\right\} 
\nonumber \\
 & & \mbox{} +\frac{8\pi}{9}\frac{\alpha_s}{Mm}\delta^{(3)}(r)
\vec\sigma_Q\cdot\vec\sigma_{\bar q}
+\frac{\alpha_s}{3Mm r^3}S_{12} \nonumber
\\
 & & \mbox{} +\frac{2\alpha_s}{3 r^3}
\left\{\frac{M^2+m^2}{2M^2m^2}+\frac{2}{Mm}\right\}
\vec S\cdot \vec L \nonumber \\ 
&&-{\alpha_s\over 6 r^3}{M^2-m^2\over M^2m^2}(\vec\sigma_Q
-\vec \sigma_{\bar q})\cdot \vec L
+ \frac{\alpha_s}{4 r^5 M^2m^2}Q_{12}. \label{gluon}
\end{eqnarray}
Here the mass coefficient $m_2$ is defined in~(\ref{mass}).
Note that without 
the square root factors in~(\ref{qpot}) the numerical
coefficients in the last two terms in the first bracket on the r.h.s.
would be $-\pi$ and $1$ instead of $-3\pi/2$ and $1/2$ respectively.

For light quarkonia the static gluon exchange interaction~(\ref{gluon}) 
has only qualitative value in view of the slow convergence
of the asymptotic expansion in $v/c$. Moreover the quadratic
spin-orbit interaction is ill-behaved in this limit. A 
more realistic version of this interaction is obtained by 
replacing the main $1/r$ (and the accompanying $\delta$ function
term) by the corresponding unapproximated
form:
\begin{equation}
-\frac{4}{3}\frac{\alpha_s}{r}\rightarrow
-\frac{4}{3}\frac{f_0(r)}{r}. \label{rep}
\end{equation}
Here the function $f_0(r)$ is defined as
\begin{equation}
f_0(r)=\frac{2}{\pi}\int_{0}^{\infty}dk\frac{\sin(kr)}{k}
\frac{M}{e_Q}\frac{m}{e_{\bar q}}
\left(\frac{M+m}{e_Q+e_{\bar q}}\right)\alpha_s(k^2).
\end{equation}
The factors $e_Q$ and $e_{\bar q}$ are defined
as
\begin{equation}
e_Q=\sqrt{M^2+\frac{k^2}{4}},\quad
e_{\bar q}=\sqrt{m^2+\frac{k^2}{4}}.
\end{equation}

Here $\alpha_s(k^2)$ is the running coupling constant of
QCD, for which we use the 
screened parametrization \cite{Matti}

\begin{equation}
\alpha_s(k^2)=\frac{12\pi}{27}\frac{1} 
{\ln [(k^2+4m_g^2)/\Lambda_0^2]}.\label{zzz}
\end{equation}

The corresponding modification of the $\vec p\,^2/r$ term
in~(\ref{gluon}) is obtained by the replacement
\begin{equation}
-\frac{4}{3}\alpha_s\frac{\vec p\,^2}{2 Mm r}
\rightarrow -\frac{4}{3}\frac{f_2(r)}{r}\vec p\,^2,
\end{equation}
where
\begin{eqnarray}
f_2(r)&=&\frac{2}{\pi}\int_0^\infty dk \frac{\sin(kr)}{k}
\left(\frac{e_Q+M}{2 e_Q}\right)\left(\frac{e_{\bar q}+m}{2 e_{\bar q} }\right)
\left(\frac{M+m}{e_Q + e_{\bar q}}\right)
\nonumber \\
&&\left\{\frac{4}{(e_Q+M)(e_{\bar q}+m)}-\frac{1}{2e_Qe_{\bar q}}
+\frac{1}{(e_Q+M)^2}
\left[1-\frac{M(e_Q+M)}{2 e_Q^{2}}\right]\right.
\nonumber\\
&&+\left.\frac{1}{(e_{\bar q}+m)^2}\left[1-\frac{m(e_{\bar q}+m)}
{2e_{\bar q}^2}\right]\right\}\alpha_s(k^2). \label{f2}
\end{eqnarray}
It is necessary to treat this term exactly in the
solution of the wave equation and not as a first order
perturbation.
The unapproximated version of the spin-spin component
of the gluon exchange interaction is introduced by
the replacement of the delta function as
\begin{equation}
\alpha_s \delta^{(3)}(r)\rightarrow
\frac{1}{2\pi^2 r}\int_0^\infty dk k \sin(kr)
\left(\frac{M+m}{e_Q+e_{\bar q}}\right)\frac{Mm}{e_Q e_{\bar q}}\alpha_s(k^2).
\end{equation}
The unapproximated forms for the tensor and spin-orbit interactions 
in~(\ref{gluon}) are
\begin{equation}
V_G(T)=\frac{2}{9 \pi} S_{12}\int_0^\infty dk k^2 j_2(kr)
\left(\frac{M+m}{e_Q +e_{\bar q}}\right)\frac{\alpha_s(k^2)}{e_Q e_{\bar q}}.
\end{equation}
\begin{eqnarray}
V_G(LS) & = & \frac{2}{3 \pi r}\vec S\cdot \vec L
\int_0^\infty dk k j_1(kr)\left(\frac{M+m}{e_Q+e_{\bar q}}\right)
\frac{e_Q+M}{e_Q}\frac{e_{\bar q}+m}{e_{\bar q}} \nonumber\\
& & \left\{\frac{1}{(e_{\bar q}+m)^2}\left[1-\frac{k^2}{4(e_Q+M)^2}\right]
+\frac{1}{(e_Q+M)^2}\left[1-\frac{k^2}{4(e_{\bar q}+m)^2}
\right]\right. \nonumber\\
& & \left. \mbox{} +\frac{4}{(e_Q+M)(e_{\bar q}+m)}\right\}\alpha_s(k^2).
\end{eqnarray}
The integrals in the spin-spin and tensor interactions
above are strongly moderated when the running coupling
strength $\alpha_s (k^2)$ is taken into account.
The quadratic spin-orbit interaction component in~(\ref{gluon}) 
may be similarly regulated by employment of the unapproximated
form
\begin{equation}
\frac{\alpha_s}{4 r^5 M^2m^2}
\rightarrow \frac{2}{3\pi r^2}\int_0^\infty dk k^2  j_2(kr)
\frac{\alpha_s(k^2)}{e_Q e_{\bar q}(e_Q+M)(e_{\bar q}+m)}
\left(\frac{M+m}{e_Q+e_{\bar q}}\right). \label{GSO2} 
\end{equation}

This form of the quadratic spin-orbit interaction has 
finite matrix elements.

The parametrization~(\ref{zzz}) takes the long distance screening
of the quark-gluon coupling into account through the gluon mass
parameter $m_g$. For this we use the value $m_g=$ 240 MeV
while for the confinement scale parameter $\Lambda_0$
we use the value 280 MeV. These values are chosen by
a fit to the heavy-light meson spectra below. They are similar
to those used in refs. \cite{Matti,Brod}.

\begin{figure}[h!]
\begin{center}
\epsfig{file=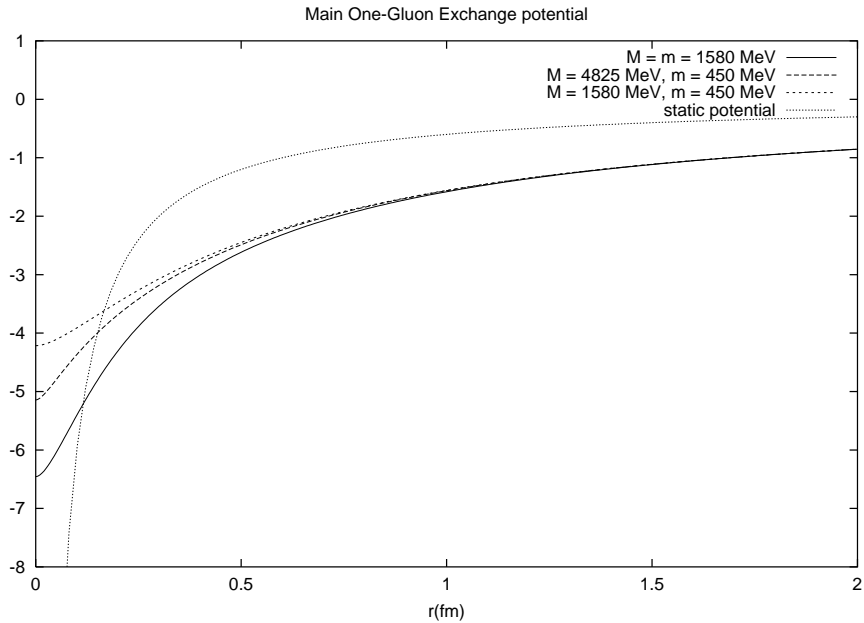, width=12cm}
\caption{The exact function $-\frac{4}{3}\frac{f_0(r)}{r}$ 
calculated for several sets of heavy and light quark
masses and compared to the static approximation for
$\alpha_s$ = 0.45} \label{func}
\end{center}
\end{figure}

The employment of the screened running quark-gluon 
coupling strength along with the relativistic corrections
to the gluon exchange interaction leads to significant
dampening of the gluon exchange interaction at short range.
This is shown in Fig.~\ref{func}, where the unapproximated 
values for 
$f_0(r)$ as given by the expression~(\ref{rep}) 
are plotted for the case of D and B
mesons with $M=1.58$ GeV and 4.825 GeV respectively,
and $m=$ 450 MeV. The result reveals that the 
approximation $f_0(r)=\alpha_s$ = constant is inadequate.
The parametrization~(\ref{zzz}) gives the value $\sim 0.4$ for
$\alpha_s$ at the charmonium scale, which is consistent
with the values extracted by lattice methods in refs.
\cite{Dav1,Dav2}. In the numerical results for the
spectra of the heavy-light mesons below, it proved
essential to use the unapproximated expressions
for the functions $f_0(r)$ and $f_2(r)$, the reason
being that the small mass of the light quarks
render the static approximations misleading.

\subsection{The instanton induced interaction}

\vspace{0.5cm}

Numerical QCD lattice calculations indicate that the
instanton liquid model provides an appropriate
description of the dynamics in the low energy regime
\cite{Diak,Negele1,Negele2}. The instantons induce
a pointlike interaction between heavy and light
quarks, which contributes significantly to the
hyperfine interaction in heavy-light mesons
\cite{Nowak}. 

	The instanton induced interaction for $Q\bar q$
systems contains a color dependent component, which may
be the cause of the splitting between
light pseudoscalar and vector mesons, for which the color
magnetic interaction is screened.
The form of the instanton induced interaction, which
is appropriate for heavy-light mesons has been derived
in ref.\cite{Nowak}, who considered the large $N_C$
limit, where the matrix element of the color exchange
operator is $\left<\lambda^1\cdot \lambda^2\right>/4=-N_c/2$. The
spin-dependent component of this interaction 
couples as a $T$ invariant
$\sigma_{\mu\nu}^1\sigma_{\mu\nu}^2 /2$,
which implies that that it may be viewed as a vector
meson exchange like interaction with anomalous
couplings to quarks. 

If the large $N_c$ limit is not invoked, the instanton
induced interaction for the $Q\bar q$ system takes the
form \cite{Nowak}:

\begin{eqnarray}
H_{Q\bar q}&=&\left(\frac{\Delta M_Q\Delta m_q}{2nN_c}\right)
\left(\mathtt{1}+\frac{1}{4}\lambda_Q\cdot\lambda_{\bar q}\right)\,
\delta^{(3)}(r) \nonumber \\ 
&&-\frac{1}{4}
\left(\frac{\Delta M_Q^{spin}\Delta m_{\bar q}}{2nN_c}\right)
\vec\sigma_Q\cdot\vec\sigma_{\bar q}  
\,\lambda_Q\cdot\lambda_{\bar q}\,
\delta^{(3)}(r)
\label{inst}
\end{eqnarray}

Here $n$ represents the instanton density, which 
typically is assigned the value 1 fm$^{-4}$. The mass 
parameters $\Delta M_Q$,
$\Delta m_{\bar q}$ and $\Delta M_Q^{spin}$ represent
the mass shifts of the heavy ($Q$) and light ($\bar q$)
quarks, which are caused by the instanton induced interaction.
The numerical values for these mass shifts were obtained
as $\Delta M_Q$= 86 MeV, $\Delta M_Q^{spin}$= 3 MeV
and $\Delta m_q$=420 MeV respectively in ref. \cite{Nowak}.

The expression~(\ref{inst}) represents the approximate effective 
instanton induced interaction that obtains 
when one of the quarks is much heavier than the other 
one, which is appropriate here. The Lorentz structure
of the main term in the interaction is a scalar, whereas
the spin dependent term has a tensor coupling structure.

In the large color limit the sign of the spin-independent term
in~(\ref{inst})
is negative. In that limit the instanton induced
interaction~(\ref{inst}) plays a role akin to that of gluon exchange,
with matrix elements of similar magnitude \cite{Nowak}.
Without invocation of that limit, the spin-dependent
term of the interaction~(\ref{inst})
is not sufficiently strong to explain e.g. all of the difference
between the $D^*$ and $D$ meson masses. As below it is found
that the screened gluon exchange interaction described above
provides sufficient splitting strength, the numerical
results do not include the interaction~(\ref{inst}). If the gluon
mass parameter were increased, it would be necessary to
include the additional spin-dependence given by the
instanton induced interaction, however.


\section{The spectra of charmonium and the heavy-light mesons
} \label{sec_spec}

\vspace{0.5cm}

\subsection{The charmonium spectrum} \label{cc_mes}

The mass of the charm quark was fixed by a fit to the $D$ and $D_S$ meson
spectra, and then tested by direct computation of the $c\bar c$ spectrum,
without further modification of the charm quark mass.
The spectrum of charmonium is fairly well known experimentally.

The interaction model consists of the all the local
terms in the scalar linear confining interaction~(\ref{conf}),(\ref{c4})
and the gluon exchange interaction, with full account
of the relativistic corrections and the running coupling
strength as described in section~(\ref{oge}) above.

The calculated spectrum is compared to the experimental
spectrum in Table~\ref{ccstate}. The parameters in the interaction model
are listed in Table~\ref{hparam}. 
The value for the charm quark mass (1580 MeV) was chosen 
to be the same as that used for the
$D$ and $D_S$-mesons below. With the parameter values
listed in Table \ref{hparam} we are able to accurately reproduce the
experimentally determined $J/\Psi$-$\eta_c$ splitting, but at the price of
30-50 MeV overpredictions of the excited states. This effect is similar to
that noted in \cite{Gross}. Better agreement with
experiment can be achieved by lowering the confining string constant to
960 MeV/fm and raising the gluon mass $m_g$ by 20 MeV to 260 MeV. In this
case the excited states agree fairly well with experiment, while the 
$J/\Psi$-$\eta_c$ splitting is underpredicted by 15 MeV. A lower charm
quark mass for the charmonium system would also improve the spectra,
but as the charmonium spectrum is calculated primarily for testing
purposes 
we keep the charm quark mass equal to that used for the $D$
and $D_S$-mesons. We thus conclude that our test results for charmonium
are similar to the spectrum obtained in \cite{Gross}.

\begin{table}[h!]
\begin{center}
\begin{tabular}{c|c c c}
   	& $c\bar c$	& $D_S$		& $D$ \\ \hline \hline
c  	& 1120 MeV/fm	& 1120 MeV/fm	& 1120 MeV/fm 	\\
b  	& -50 MeV	& -260 MeV	& -320 MeV 	\\
$\Lambda_{0}$	& 280 MeV & 280 MeV & 280 MeV		\\
$m_g$	& 240 MeV	& 240 MeV	&  240 MeV	\\ \hline
$m_c$	& 1580 MeV	& 1580 MeV	&  1580 MeV	\\
$m_s$	&  -	& 560 MeV	&  - 	\\
$m_{u,d}$   	& - 		&  -	& 450 MeV	\\
\end{tabular}
\caption{Model parameters used for the charmonium, $D$ and
$D_S$ meson spectra.} \label{hparam}
\end{center}
\end{table}

The quality of the spectrum is
similar to that obtained with an effective interaction
constructed by means of lattice methods \cite{Bali},
as well as by completely nonrelativistic phenomenology
\cite{Quigg,Timo}. The numerical values for the calculated
energies of the $c\bar c$ are listed in Table ~\ref{ccstate}, along 
with the experimental values \cite{Caso}.

\begin{table}[h!]
\begin{center}
\begin{tabular}{c||c|c|}
& Predicted States (MeV) & Experimental states (MeV)  \\ \hline\hline
$1 ^{1}S_{0}$ & 2975 & 2979.8  $\pm$ 2.1	\\
$1 ^{3}S_{1}$ & 3088 & 3096.88 $\pm$ 0.04	\\
$2 ^{1}S_{0}$ & 3682 & 				\\
$2 ^{3}S_{1}$ & 3736 & 3686.00 $\pm$ 0.09	\\ \hline
$1 ^{1}P_{1}$ &	3518 & 				\\
$1 ^{3}P_{0}$ &	3450 & 3417.3  $\pm$ 2.8	\\
$1 ^{3}P_{1}$ &	3519 & 3510.53 $\pm$ 0.12	\\
$1 ^{3}P_{2}$ &	3580 & 3556.17 $\pm$ 0.13	\\ \hline
\end{tabular}
\caption{Calculated and experimental charmonium states.} \label{ccstate}
\end{center}
\end{table}

\newpage

\subsection{The $D$ and $D_s$ -meson spectra} \label{d_mes}

In the calculation of the spectra of the $D$ and $D_s$ mesons,
the constituent masses of the light and strange quarks
are treated as phenomenological parameters to be fitted
against the known splittings in the spectrum. While a reasonable
overall description of the $D$ meson spectrum is readily
obtainable with the interaction models described
above, a quantitative description of e.g. the spin-orbit
splitting of the P-shell $D_j$ ($j=0,1,2$) $D$ meson
resonances demands a real effort.

The main difficulty with the spin orbit splittings may be
ascribed to the small mass of the light quarks, which makes
the matrix elements of the spin-orbit components of both
the confining and hyperfine interactions large. That is
the reason for taking into account the relativistic
damping of these interaction components that is described
in section~\ref{int_ham}.

\begin{table}[h!]
\begin{center}
\begin{tabular}{c||c|c c|}
& $M_{u,d}$=450 MeV & Exp($D^0$) & Exp($D^{\pm}$) \\ \hline\hline
$1 ^{1}S_{0}$ & 1874 & 1864.6$\pm$0.5 & 1869.3$\pm$0.5 \\
$1 ^{3}S_{1}$ & 2006 & 2006.7$\pm$0.5 & 2010.0$\pm$0.5 \\
$2 ^{1}S_{0}$ & 2540 &		&		\\
$2 ^{3}S_{1}$ & 2601 & 2637 ?	& 2637 ?	\\
$3 ^{1}S_{0}$ & 2904 &		&		\\
$3 ^{3}S_{1}$ & 2947 &		&		\\
$4 ^{1}S_{0}$ & 3175 &		&		\\
$4 ^{3}S_{1}$ & 3208 &		&		\\ \hline
$1 ^{1}P_{1}$ &	2389 &		&		\\
$1 ^{3}P_{0}$ &	2341 &		&		\\
$1 ^{3}P_{1}$ &	2407 & 2422.2$\pm$1.8 &		\\
$1 ^{3}P_{2}$ &	2477 & 2458.9$\pm$2.0 & 2459 $\pm$ 4	\\
$2 ^{1}P_{1}$ &	2792 &		&		\\
$2 ^{3}P_{0}$ &	2758 &		&		\\
$2 ^{3}P_{1}$ &	2802 &		&		\\
$2 ^{3}P_{2}$ &	2860 &		&		\\
$3 ^{1}P_{1}$ &	3082 &		&		\\
$3 ^{3}P_{0}$ &	3050 &		&		\\
$3 ^{3}P_{1}$ &	3085 &		&		\\
$3 ^{3}P_{2}$ &	3142 &		&		\\ \hline
$1 ^{1}D_{2}$ &	2689 &		&		\\
$1 ^{3}D_{1}$ &	2750 &		&		\\
$1 ^{3}D_{2}$ &	2727 &		&		\\
$1 ^{3}D_{3}$ &	2688 &		&		\\ 
$2 ^{1}D_{2}$ &	2997 &		&		\\
$2 ^{3}D_{1}$ &	3052 &		&		\\
$2 ^{3}D_{2}$ &	3029 &		&		\\
$2 ^{3}D_{3}$ &	2999 &		&		\\ \hline
\end{tabular}
\caption{Calculated and experimental $D$-meson states.}
\label{tabd_mes}
\end{center}
\end{table}

\newpage
\begin{figure}[h!]
\begin{center}
\epsfig{file=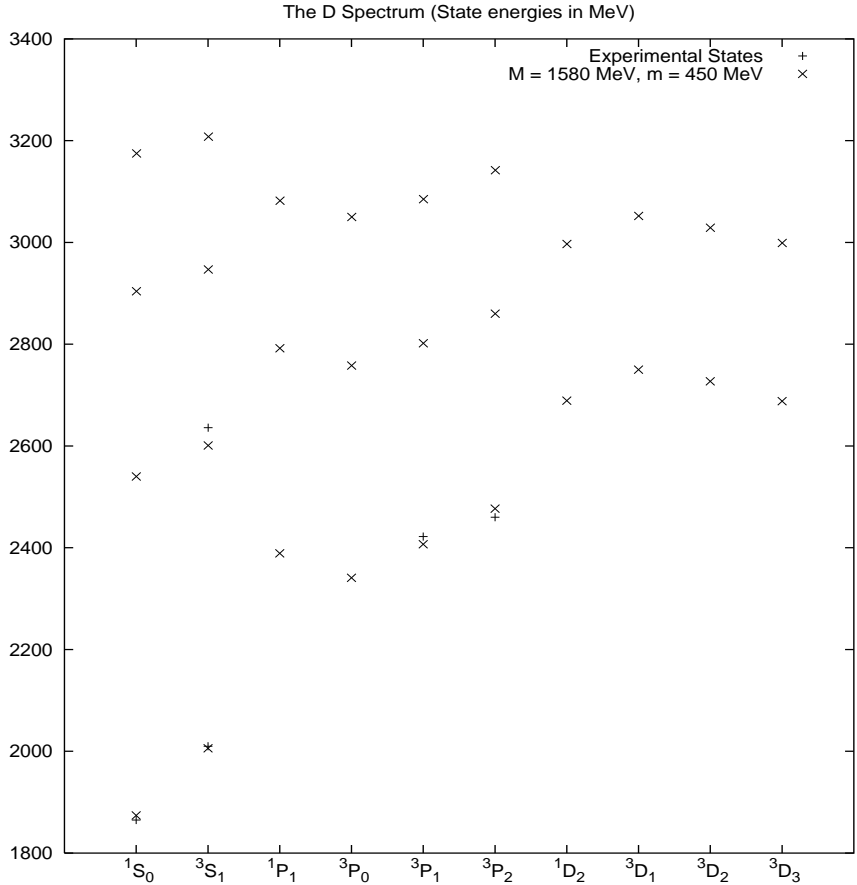, width=12cm, height=12cm}
\caption{Calculated and experimental $D$-meson states.} \label{D_fig}
\end{center}
\end{figure}

The spectrum of the $D$ meson that is obtained with the
scalar linear confining interaction and with the
quark masses $m_u=m_d=$ 450 MeV and $m_c=$ 1580 MeV is
shown in Fig.~\ref{D_fig}. The light quark mass here is about 100 MeV
larger than the typical values employed in nonrelativistic
phenomenology. Reducing this mass further in the calculation
would lead to an unrealistically large spin-orbit
splitting of the $P-$ states, while giving an unrealistically small
\mbox{1S $\rightarrow$ 2S} spacing. The interaction parameters
used in the calculation are listed in Table~\ref{hparam}. 
 
The calculated energies of the
$D-$meson states are listed in 
Table~\ref{tabd_mes} along with the known empirical
energies \cite{Caso}.

The calculated $D$ meson spectrum has the correct $D^* -D$
ground state splitting and only slightly underpredicts
the possible $D$ meson excitation \cite{Bourd} at 2637 MeV.
The $P$-shell states around 2400 MeV are also
well reproduced.  

\begin{figure}[h!]
\begin{center}
\epsfig{file=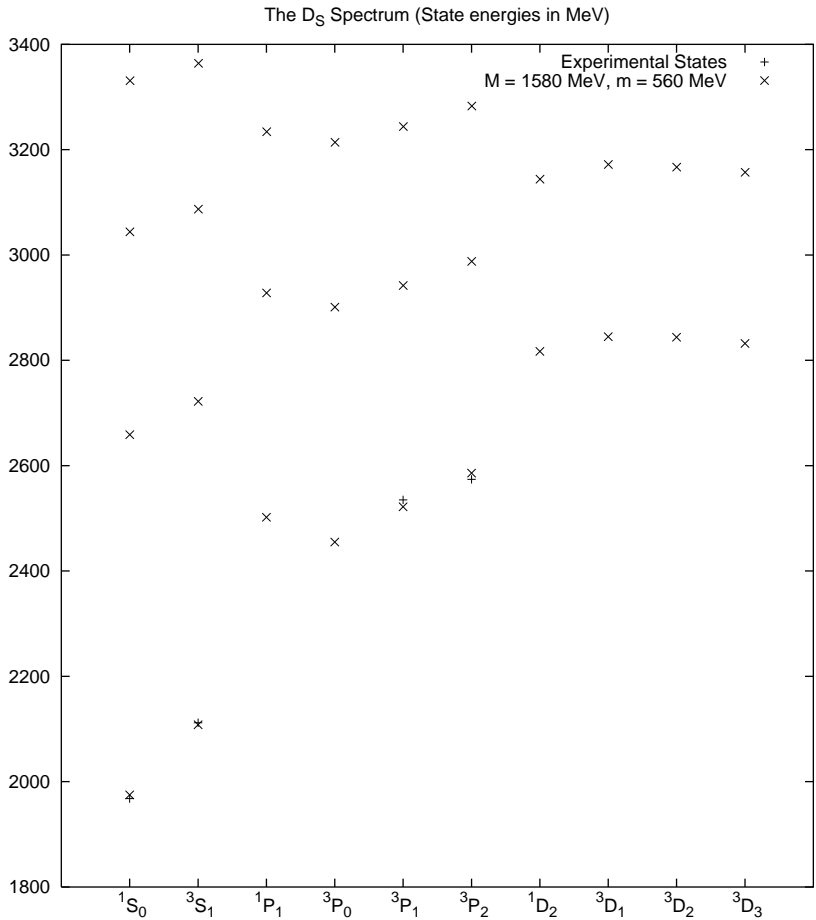}
\caption{Calculated and experimental $D_s$ meson states.} \label{Ds_fig}
\end{center}
\end{figure}

The calculated energies of the $D_s$ meson states are
shown in Fig.~\ref{Ds_fig} and are listed in Table~\ref{tabds_mes}. As
in the
case of the $D$ meson a satisfactory description of the
still very incompletely known experimental spectrum
is only achievable with a fairly large value for the
constituent mass of the $s$ quark ($m_s=$ 560 MeV). 

The overall structure of the $D_s$ meson spectrum in
Fig.~\ref{Ds_fig} is similar to that of the $D$ meson spectrum
in Fig.~\ref{D_fig}. The ground state splitting is given correctly,
as are the known $P$-state excitations.

The calculated $D$ meson spectrum is similar
to that obtained by the Gross reduction of the
Bethe-Salpeter equation in ref.\cite{Wally}. The
shell spacing at increasing energy of the latter
is somewhat wider, which may reflect a difference
between the different quasipotential reductions,
but also the employment of static potentials in
ref.\cite{Wally}.

\begin{table}[h!]
\begin{center}
\begin{tabular}{c||c|c|}
& $M_s$=560 MeV & Experimental states (MeV)	\\ \hline\hline
$1 ^{1}S_{0}$ & 1975 & 1968.5 $\pm$ 0.6	\\
$1 ^{3}S_{1}$ & 2108 & 2112.4 $\pm$ 0.7	\\
$2 ^{1}S_{0}$ & 2659 &	\\
$2 ^{3}S_{1}$ & 2722 &	\\
$3 ^{1}S_{0}$ & 3044 &	\\
$3 ^{3}S_{1}$ & 3087 &	\\
$4 ^{1}S_{0}$ & 3331 &	\\
$4 ^{3}S_{1}$ & 3364 &	\\ \hline
$1 ^{1}P_{1}$ &	2502 &	\\
$1 ^{3}P_{0}$ &	2455 &	\\
$1 ^{3}P_{1}$ &	2522 & 2535.35 $\pm$ 0.34 $\pm$ 0.5	\\
$1 ^{3}P_{2}$ &	2586 & 2573.5 $\pm$ 1.7  \\
$2 ^{1}P_{1}$ &	2928 &	\\
$2 ^{3}P_{0}$ &	2901 &	\\
$2 ^{3}P_{1}$ &	2942 &	\\
$2 ^{3}P_{2}$ &	2988 &	\\
$3 ^{1}P_{1}$ &	3234 &	\\
$3 ^{3}P_{0}$ &	3214 &	\\
$3 ^{3}P_{1}$ &	3244 &	\\
$3 ^{3}P_{2}$ &	3283 &	\\ \hline
$1 ^{1}D_{2}$ &	2838 &	\\
$1 ^{3}D_{1}$ &	2845 &	\\
$1 ^{3}D_{2}$ &	2856 &	\\
$1 ^{3}D_{3}$ &	2857 &	\\
$2 ^{1}D_{2}$ &	3144 &	\\
$2 ^{3}D_{1}$ &	3172 &	\\
$2 ^{3}D_{2}$ &	3167 &	\\
$2 ^{3}D_{3}$ &	3157 &	\\ \hline
\end{tabular}
\caption{Calculated and experimental $D_s$-meson states.} \label{tabds_mes}
\end{center}
\end{table}

\newpage

\subsection{The $B$ and $B_s$ -meson spectra} \label{b_mes}

The empirical knowledge of the spectra of the $B$ and
$B_s$ mesons is still very incomplete. Only the ground
state pseudoscalar and vector meson state energies are
known with certainty. In addition one 
orbital excitation of the $B$ meson is known at 5700 MeV,
which presumably belongs to the $P-$shell.

The only additional parameter required to calculate the
spectra of the bottom mesons with the interaction
models described above is the $b$ quark mass. For this
we choose the value $m_b$=4825 MeV, with which value
the pseudoscalar-vector splitting of the bottom and
strange bottom mesons is obtained correctly.

\begin{figure}[h!]
\begin{center}
\epsfig{file=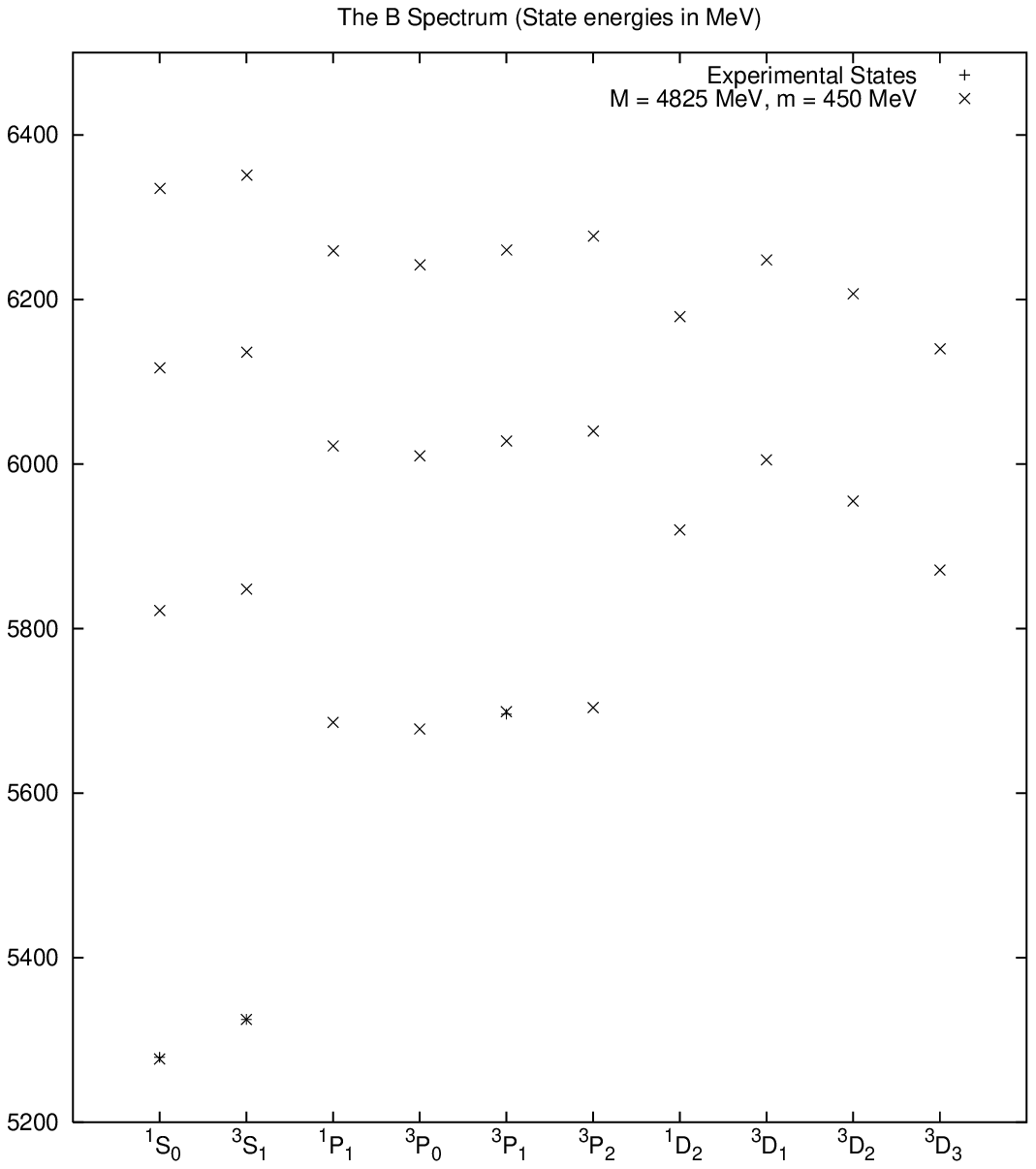}
\caption{Calculated and experimental $B$-meson spectra.} \label{B_fig}
\end{center}
\end{figure}

\begin{figure}[h!]
\begin{center}
\epsfig{file=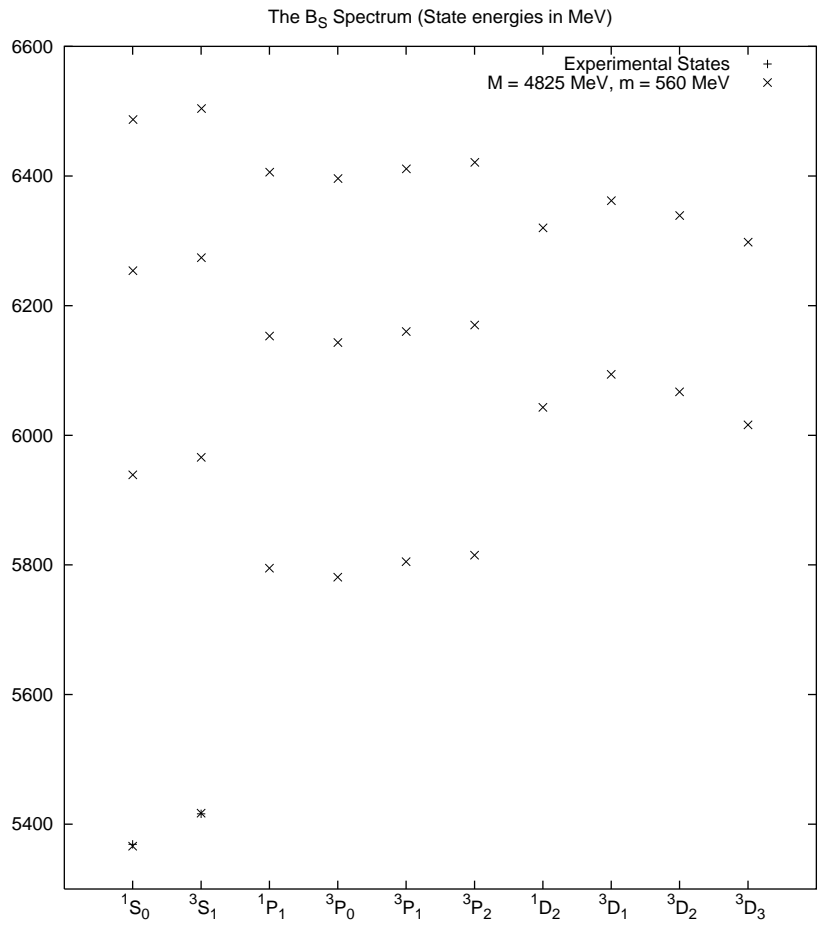}
\caption{Calculated and experimental $B_s$-meson spectra.} \label{Bs_fig}
\end{center}
\end{figure}

The calculated $B$ and $B_s$ meson spectra are shown
in Figs.~\ref{B_fig} and~\ref{Bs_fig}. The energies of these states are
also listed in Tables ~\ref{tabb_mes} and~\ref{tabbs_mes}, along with the
empirical
values that are taken from ref.\cite{Caso}. The
orbital excitation of the $B$ meson at 5700 MeV is
obtained correctly, under the assumption that it
corresponds to a $j=1$ $P$-shell state.

The overall features of the calculated bottom meson
spectra are similar to those obtained with the
Gross reduction \cite{Gross} of the Bethe-Salpeter equation
in ref.\cite{Wally}, although as in the case of the
charm mesons, we obtain somewhat smaller shell
spacings at higher excitation.

\begin{table}[h!]
\begin{center}
\begin{tabular}{c|c c}
	& $B_s$		& $B$ 		\\ \hline \hline
c	& 1120 MeV/fm	& 1120 MeV/fm	\\
b	& -185 MeV	& -250 MeV	\\
$\Lambda_{0}$	& 280 MeV & 280 MeV	\\
$m_g$	& 240 MeV	& 240 MeV	\\ \hline
$m_b$	& 4825 MeV	& 4825 MeV	\\ 
$m_s$	& 560 MeV	& -		\\
$m_{u,d}$& -		& 450 MeV	\\
\end{tabular}
\caption{Parameter values used in the calculation
of the $B_S$ and $B$-meson spectra.} \label{lparam}
\end{center}
\end{table}

\begin{table}[h!]
\begin{center}
\begin{tabular}{c||c|c c|}
State & $M_{u,d}$=450 MeV & Exp($B^0$) & Exp($B^{\pm}$) \\ \hline\hline
$1 ^{1}S_{0}$ & 5277 & 5279.2$\pm$1.8 & 5278.9$\pm$1.8 \\
$1 ^{3}S_{1}$ & 5325 & 5324.9$\pm$1.8 & 	\\
$2 ^{1}S_{0}$ & 5822 &		&		\\
$2 ^{3}S_{1}$ & 5848 & 		& 		\\
$3 ^{1}S_{0}$ & 6117 &		&		\\
$3 ^{3}S_{1}$ & 6136 &		&		\\
$4 ^{1}S_{0}$ & 6335 &		&		\\
$4 ^{3}S_{1}$ & 6351 &		&		\\ \hline
$1 ^{1}P_{1}$ &	5686 &		&		\\
$1 ^{3}P_{0}$ &	5678 &		&		\\
$1 ^{3}P_{1}$ &	5699 & 5697$\pm$9 &		\\
$1 ^{3}P_{2}$ &	5704 & 		& 		\\
$2 ^{1}P_{1}$ &	6022 &		&		\\
$2 ^{3}P_{0}$ &	6010 &		&		\\
$2 ^{3}P_{1}$ &	6028 &		&		\\
$2 ^{3}P_{2}$ &	6040 &		&		\\
$3 ^{1}P_{1}$ &	6259 &		&		\\
$3 ^{3}P_{0}$ &	6242 &		&		\\
$3 ^{3}P_{1}$ &	6260 &		&		\\
$3 ^{3}P_{2}$ &	6277 &		&		\\ \hline
$1 ^{1}D_{2}$ &	5920 &		&		\\
$1 ^{3}D_{1}$ &	6005 &		&		\\
$1 ^{3}D_{2}$ &	5955 &		&		\\
$1 ^{3}D_{3}$ &	5871 &		&		\\ 
$2 ^{1}D_{2}$ &	6179 &		&		\\
$2 ^{3}D_{1}$ &	6248 &		&		\\
$2 ^{3}D_{2}$ &	6207 &		&		\\
$2 ^{3}D_{3}$ &	6140 &		&		\\ \hline
\end{tabular}
\caption{Calculated and experimental $B$-meson states} \label{tabb_mes}
\end{center}
\end{table}

\newpage

\begin{table}[h!]
\begin{center}
\begin{tabular}{c||c|c|}
& $M_s$=560 MeV & Experimental $B_S$ states (MeV) \\ \hline\hline
$1 ^{1}S_{0}$ & 5366 & 5369.3 $\pm$ 2.0	\\
$1 ^{3}S_{1}$ & 5417 & 5416.3 $\pm$ 3.3	\\
$2 ^{1}S_{0}$ & 5939 &	\\
$2 ^{3}S_{1}$ & 5966 &	\\
$3 ^{1}S_{0}$ & 6254 &	\\
$3 ^{3}S_{1}$ & 6274 &	\\
$4 ^{1}S_{0}$ & 6487 &	\\
$4 ^{3}S_{1}$ & 6504 &	\\ \hline
$1 ^{1}P_{1}$ &	5795 &	\\
$1 ^{3}P_{0}$ &	5781 &	\\
$1 ^{3}P_{1}$ &	5805 &  \\
$1 ^{3}P_{2}$ &	5815 &  \\
$2 ^{1}P_{1}$ &	6153 &	\\
$2 ^{3}P_{0}$ &	6143 &	\\
$2 ^{3}P_{1}$ &	6160 &	\\
$2 ^{3}P_{2}$ &	6170 &	\\
$3 ^{1}P_{1}$ &	6406 &	\\
$3 ^{3}P_{0}$ &	6396 &	\\
$3 ^{3}P_{1}$ &	6411 &	\\
$3 ^{3}P_{2}$ &	6421 &	\\ \hline
$1 ^{1}D_{2}$ &	6043 &	\\
$1 ^{3}D_{1}$ &	6094 &	\\
$1 ^{3}D_{2}$ &	6067 &	\\
$1 ^{3}D_{3}$ &	6016 &	\\
$2 ^{1}D_{2}$ &	6320 &	\\
$2 ^{3}D_{1}$ &	6362 &	\\
$2 ^{3}D_{2}$ &	6339 &	\\
$2 ^{3}D_{3}$ &	6298 &	\\ \hline
\end{tabular}
\caption{Calculated and experimental $B_s$-meson states.} \label{tabbs_mes}
\end{center}
\end{table}

\newpage

\newpage

\newpage

\section{Quark current operators} \label{sec_curr}

\vspace{0.5cm}

\subsection{Single quark currents}

In order to calculate the  M1 transitions of the $Q\bar q$ systems
we employ the full Dirac structure of the single quark
current operators. This was shown to be necessary in 
ref.\cite{Timo} for a satisfactory description of the
corresponding M1 decays of heavy quarkonia.
The current density operators of
the constituent light quarks and the $c$ 
and $b$ quarks are then

\begin{equation}
\left<p'|\vec \jmath\,(0)|p\right>=
iQ[\bar u(\vec p\,')\vec \gamma u(\vec{p}\,)].
\end{equation}
Here $Q$ is the quark charge operator, which takes the values $\pm 2e/3$
and $\pm e/3$. With canonical boosts for the quark spinors $u,\bar u$,
this current operator leads to the magnetic moment operator
\begin{equation}
\vec \mu=Q\left (\frac{m_p}{m_Q}\right)\frac{\vec \sigma}{\sqrt{1+\vec
v\,^2}}\left\{1-\frac{1}{3}\left(1-\frac{1}{\sqrt{1+\vec v\,^2}}\right)
\right\}\mu_N,
\end{equation}
where $\vec v$ is the quark velocity: $\vec v=(\vec p\,'
+\vec p\,)/2m_Q$,
$m_p$ is the proton mass and $\mu_N$ is the nuclear magneton.

For the $D^{\pm}$ mesons the spin flip part of the combination of
the magnetic moments of the charm quark and the light antiquarks is

\begin{eqnarray}
\vec\mu_{flip}(D^{\pm}) & = &\pm \frac{1}{6}m_p(\vec \sigma_c-\vec
\sigma_{\bar d})  
\left\{\frac{2}{m_c}\frac{1}{\sqrt{1+\vec v\,_c^2}}\left[1-\frac{1}{3}
\left(1-\frac{1}{\sqrt{1+\vec v\,_c^2}}\right)\right]\right. \nonumber \\
& & \left.  
-\frac{1}{m_{\bar d}}\frac{1}{\sqrt{1+\vec v\,_{\bar d}^2}}
\left[1-\frac{1}{3}\left(1-\frac{1}{\sqrt{1+\vec v\,_{\bar d}^2}}
\right)\right]\right\}\mu_N. \label{mu_Dpm}
\end{eqnarray}

For the $D^0$, $\bar D^0$ mesons the corresponding operator is

\begin{eqnarray}
\vec \mu_{flip}\left(D^0,\bar D^0\right) & = & \pm \frac{1}{3}m_p\left(
\sigma_c-\sigma_{\bar
u}\right)\left\{\frac{1}{m_c}\frac{1}{\sqrt{1+
\vec v\,_c^2}}\left[1-\frac{1}{3}
\left(1-\frac{1}{\sqrt{1+\vec v\,_c^2}}\right)\right]\right. \nonumber \\
& & \left. \mbox{} +\frac{1}{m_{\bar u}}\frac{1}{\sqrt{1+
\vec v\,_{\bar u}^2}}
\left[1-\frac{1}{3}\left(1-\frac{1}{\sqrt{1+\vec v\,_{\bar u}^2}} \right)
\right]\right\}\mu_N. \label{mu_D0}
\end{eqnarray}

Here $m$ is the mass of the light antiquark, and $\vec v_c$ and $\vec
v_{\bar q}$ denote the velocity operators of the charm quark and the
light antiquark respectively. The spin-flip magnetic moment operator for
the $D^\pm_s$ meson may be obtained from the corresponding expression
for the $D^\pm$ mesons by replacing $\vec \sigma_{\bar d}$, $m_{\bar d}$
and $\vec v_{\bar d}$ in ~(\ref{mu_Dpm}) 
by $\vec \sigma_{\bar s}$, $m_{\bar s}$ and
$\vec v_{\bar s}$ respectively.

The spin-flip magnetic moment operators for the bottom mesons may
be constructed by similar replacements. For the $B^\pm$ mesons
these operators are obtained from those for the $D^\pm$~(\ref{mu_Dpm})
by replacement of the $c$ quark mass and
velocity by the corresponding $u$ quark mass and velocities and by
replacement of the $d$ quark mass and velocity by the $b$ quark
mass and velocity.

The spin-flip magnetic moment operator for the $B^0$ and
$\bar B^0$ mesons is obtained from that for the neutral
$D$ mesons~(\ref{mu_D0}) by multiplying the operator by $-1/2$
and by replacing the $c$ and $\bar u$ quark masses and
velocities by the corresponding masses and velocities
for the $d$ and $\bar b$ quark. A further replacement
of the $d$ quark mass and velocity by the corresponding
$s$ quark mass and velocity yields the spin-flip
magnetic moment operator for the $B_s$ mesons.

Note that when matrix elements of these magnetic moment operators are
evaluated with wave functions that are solutions to the
Blankenbecler-Sugar equation~(\ref{wave}), these 
operators should be multiplied
by the factor

\begin{equation}
f_{BS}=\frac{M+m}{M\sqrt{1+v_Q^2}+m\sqrt{1+v_{\bar q}^2}}.
\label{bbfact}
\end{equation}

\subsection{Two-quark currents}

\begin{figure}[h!]
\begin{center}
\begin{picture} (370,220)

\put(75,5){$Q$}
\put(80,20){\line(0,1){100}}
\put(150,20){\line(0,1){180}}
\put(145,5){$\bar q$}
\put(80,120){\line(-1,-1){40}}
\put(40,80){\line(0,1){120}}

\multiput(82.5,120)(10,0){7}{\line(1,0){5}}
\multiput(-2.5,80)(10,0){5}{\oval(5,5)[t]}
\multiput(-7.5,80)(10,0){5}{\oval(5,5)[b]}
\put(10,90){$\gamma$}

\put(245,5){$\bar q$}
\put(250,20){\line(0,1){100}}
\put(360,20){\line(0,1){180}}
\put(355,5){$Q$}
\put(250,120){\line(1,-1){40}}
\put(290,80){\line(0,1){120}}

\multiput(292.5,80)(10,0){7}{\line(1,0){5}}
\multiput(207.5,120)(10,0){5}{\oval(5,5)[t]}
\multiput(202.5,120)(10,0){5}{\oval(5,5)[b]}
\put(220,130){$\gamma$}

\end{picture}
\end{center}
\caption{Exchange current operators associated with the effective
scalar confining and gluon exchange interactions, with
intermediate virtual $q\bar q$ excitations.} \label{fey}
\end{figure}
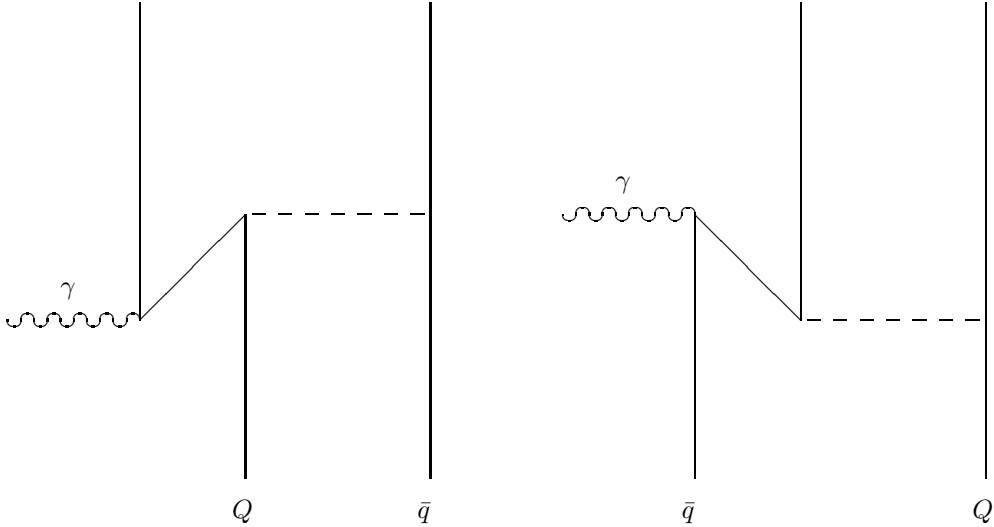

Both the scalar confining interaction and the hyperfine interaction may
excite virtual quark-antiquark states, which are deexcited by the
electromagnetic field. This process is illustrated by the Feynman
diagrams in Fig.\ref{fey} . These operators may be derived by the methods
described in ref.\cite{Timo}.

For the $D^\pm$ system the two-quark current that is implied by the
scalar confining interaction to lowest order in $v/c$ takes the form
\begin{equation}
\vec{\jmath}_2(C)=\mp ie c r\frac{m_p}{12}\left(\frac{2}{m_c^2}-\frac{1}{m_{\bar
d}^2}\right)(\vec \sigma_c-\vec \sigma_{\bar d})\times\vec q.
\end{equation}
This current operator may be viewed as a direct renormalization of the
corresponding spin-flip combination of the single $c$ and $\bar d$
current operators:
\begin{equation}
\vec{\jmath}_{c\bar d}=\pm ie \frac{m_p}{12}\left(\frac{2}{m_c^2}
-\frac{1}{m_{\bar
d}^2}\right)\left(\vec \sigma_c-\vec \sigma_{\bar d}\right)\times\vec q.
\end{equation}

The two-quark magnetic moment operator that is implied by this current
operator is then accordingly

\begin{equation}
\vec \mu_2(C)=\mp cr\frac{m_p}{6}\left\{\frac{2}{m_c^2}-\frac{1}{m_{\bar
d}^2}\right\}\left(\sigma_c-\sigma_{\bar d}\right)\mu_N. \label{mu_con_Dpm}
\end{equation}

For the $D^0,\bar D^0$ mesons the corresponding two-quark magnetic
moment is then

\begin{equation}
\vec \mu_2(C)=\mp cr\frac{m_p}{3}\left\{\frac{1}{m_c^2}-\frac{1}{m_{\bar
u}^2}\right\}\left(\sigma_c-\sigma_{\bar u}\right)\mu_N.
\end{equation}
The exchange magnetic moment operator for the $D^\pm_s$ mesons is
obtained from that for the $D^\pm$ mesons~(\ref{mu_con_Dpm}), by 
replacing $\vec\sigma_{\bar d}$ and $m_{\bar d}$ 
by $\vec \sigma_{\bar s}$ and $m_{\bar s}$ respectively.

The general rule is that the exchange current that is
associated with the confining interaction has opposite
sign to the corresponding combination of single quark
current operators, and may be obtained directly from the
former by multiplication by the factor $-cr$ and by
squaring the quark masses in the denominators. 
The corresponding operators for the bottom mesons are
readily constructed with the help of this rule.
Note that the constant $b$, which is subtracted from the
Hamiltonian, is viewed as an approximation to the velocity
dependent term $-3/2\, \vec p\,^2/m_2^2$ in the confining
interaction~(\ref{conf}). As that term may be viewed as the
origin for the two-quark interaction current that is
associated with the scalar confining interaction through the
continuity equation, it should not be subtracted from the
$cr$ term in the two-body current. 

The gluon exchange interaction~(\ref{gluon}) also 
implies a two-quark exchange
current operator. 
The matrix elements of this two-body current are however significantly
weaker than those of the exchange current operator
that is associated with the scalar confining operator \cite{Timo}.
Because of the cancellations between the matrix
elements of the single quark current operators
and those of the exchange current operator that is
associated with the confining interaction, the gluon
exchange current operator does give a significant
contribution to the calculated M1 transition rates
between the ground state heavy-light mesons.

For the $D^\pm$ mesons this two-quark current, to
lowest order in the inverse quark masses, has the form

\begin{eqnarray}
\vec{\jmath}_2(G)\left[D^\pm\right] & = 
& \pm e\frac{8\pi }{9}
\left\{
\frac{2 \alpha_s(k^2_{\bar d})}{3 k_{\bar d}^2}
\left[\frac{\vec p_{\bar d}+\vec p_{\bar d}\,'}{m_cm_{\bar
d}}\right.\right. \nonumber \\
& & \mbox{} +i\left(\frac{\vec \sigma_c}{m_c^2}+\frac{\vec \sigma_{\bar d}}
{m_cm_{\bar d}}\right)\times \vec k_{\bar d} \nonumber \\
& & \left. \mbox{} +\frac{\alpha_s(k_c^2)}
{3k_c^2}\left[\frac{\vec p_c+\vec p_c\,'}
{m_c m_{\bar d}}+i\left(\frac{\vec \sigma_c}{m_c m_{\bar d}}
+\frac{\vec \sigma_{\bar d}}{m_{\bar d}^2}\right)\times
\vec k_c\right]\right\}.
\end{eqnarray}

Here the momentum operators $\vec k_{\bar d}$ and $\vec k_c$ denote the
fractional momenta imparted to the $\bar d$ and $c$ (or $d$ and $\bar
c$) quarks respectively.

The spin-flip component of the magnetic moment operator, that is
obtained from this current operator, is then

\begin{eqnarray}
\vec\mu_2(G)\left[D^\pm\right] & = & \pm\frac{f_0(r)}{27}\frac{m_p}
{r}\left(\frac{2}{m_c^2}-\frac{1}{m_c m_{\bar d}}-\frac{1}{m_{\bar
d}^2}\right) \nonumber \\
& & \mbox{} \left(\vec \sigma_c-\vec \sigma_{\bar d}\right)\mu_N.
\label{mu_ex_Dpm}
\end{eqnarray}

In the case of the $D^0$ and $\bar D^0$ mesons the gluon exchange
current operator takes the form

\begin{eqnarray}
\vec{\jmath}_2(G)\left[D^0,\bar D^0\right] & = & \pm
e\frac{16\pi}
{27}\left\{\frac{\alpha_s(k_{\bar u}^2)}
{k_{\bar u}^2}\left[\frac{\vec p_{\bar u}\,'+\vec p_{\bar
u}}{m_cm_{\bar u}}\right.\right. \nonumber \\
& & \mbox{} + i\left(\frac{\vec \sigma_c}{m_c^2}+
\frac{\vec \sigma_{\bar u}}
{m_cm_{\bar u}}\right)\times \vec k_{\bar u} \nonumber \\
& & \left.\mbox{} -\frac{\alpha_s(k_c^2)}{k_c^2}\left[
\frac{\vec p_c\,'+\vec p_c}{m_cm_{\bar
u}}+i\left(\frac{\vec \sigma_c}{m_cm_{\bar u}}+\frac{\vec \sigma_{\bar u}}
{m_{\bar u}^2}\right)\times \vec k_c\right]\right\}.
\end{eqnarray}

The corresponding spin-flip magnetic moment operator is

\begin{equation}
\vec \mu_2(G)\left[D^0,\bar
D^0\right]=\pm\frac{2 f_0(r)}{27}\frac{m_p}
{r}\left(\frac{1}{m_c^2}-\frac{1}{m_{\bar u}^2}\right)\left(\vec
\sigma_c-\vec \sigma_{\bar u}\right)\mu_N.
\label{mug_D}
\end{equation}

The gluon exchange spin-flip 
magnetic moment for the $D^\pm_s$ mesons is obtained
from that for the $D^\pm$ mesons~(\ref{mu_ex_Dpm}) by replacing 
$\vec\sigma_{\bar d}$ and
$m_{\bar d}$ by $\vec \sigma_{\bar s}$ and $m_{\bar s}$ respectively.

The gluon exchange spin-flip magnetic
moment operator for the $B^-$ and $B^+$ mesons is obtained
from the corresponding expressions~(\ref{mug_D}) for the $D^0$ and
$\bar D^0$ mesons by replacing the $c$ quark mass and
spin operator by the $b$ quark mass and spin operator.
Similarly by replacing the $c$ quark mass and spin
operator in the expressions~(\ref{mu_ex_Dpm}) for the $D^+$ and
$D^-$ gluon exchange spin-flip magnetic moment 
operator by the $b$ quark mass and spin operator
one obtains the corresponding operator for the
$\bar B^0$ and $B^0$ mesons. Finally the
expressions for the gluon exchange spin-flip
magnetic moment for the $\bar B_s^0$ and $B_s^0$
mesons are obtained from those for the
$D_s^+$ and $D_s^-$ mesons by the replacement of the
$c$ quark mass and spin operator by the
$b$ quark mass and spin operator.

The exchange magnetic moments that are associated with the
confining and gluon exchange interactions above represent
relativistic corrections of second order in $v/c$, as
is implied by their origin as pair excitation currents
(Fig.~\ref{fey}). They thus appear in the same order as the
relativistic corrections to the single quark
magnetic moment operators. As relativistic corrections to
the exchange current operators are of quartic order in
$v/c$ we have not considered them here, in order to be 
consistent with the treatment of the confining
interaction without quartic terms in $v/c$, besides the
numerically insignificant quadratic spin-orbit
interaction. 

\section{M1 transitions} \label{sec_M1}

\vspace{0.5cm}

\subsection{M1 transition matrix elements for $D$ mesons}

The spin-flip component of all the magnetic moment operators derived
above may be written in the general form
\begin{equation}
\vec \mu=e{\cal M}\left(\vec \sigma_Q-\vec \sigma_{\bar q}\right).
\end{equation}
Here ${\cal M}$ is the matrix element of the orbital part of the
magnetic moment operator between the initial and final meson states. The
transition width for M1 transitions of the type $Q\bar
q(J=1)\rightarrow Q\bar q(J=0)\gamma$ may be expressed as
\begin{equation}
\Gamma=\frac{16}{3}\alpha_{em}\frac{M_f}{M_i}{\cal M}^2q^3.
\end{equation}
Here $\alpha_{em}$ is the fine structure constant and $q$ is the photon
momentum in the laboratory frame. The masses of the initial and final
meson states are denoted $M_i$ and $M_f$ respectively.

The matrix element ${\cal M}$ is defined as the coefficient of the
spin-flip operator $\vec\sigma_c-\sigma_{\bar q}$
in the expression for the magnetic moment operator.
It is formed as a sum of matrix elements of single
quark and exchange magnetic moment operators. In the case of the $D^\pm$
mesons, the matrix element between $S$-states of the single quark
operator~(\ref{mu_Dpm}) may be written in the form

\begin{eqnarray}
{\cal M}_{IA}(D^\pm) & = & \pm\frac{1}{3}
\int_{0}^{\infty} dp p^2\int_{-1}^{1}dz f_{BS}(p)
 \mbox{} \int_{0}^{\infty}dr'r'^{2}\int_{0}^{\infty}drr^2\varphi
_f(r') \nonumber \\
& & \mbox{} \left\{\frac{2}{m_c\sqrt{1+ v_c^2}}\left[1-\frac{1}{3}
\left(1-\frac{1}{\sqrt{1+ v_c^2}}\right)\right]\right. \nonumber \\
& & \left.\mbox{} -\frac{1}{m_{\bar d}\sqrt{1+
 v_{\bar d}^2}}\left[1-\frac{1}
{3}\left(1-\frac{1}{\sqrt{1+v_{\bar d}^2}}\right)\right]\right\} 
\label{mu_full_Dpm} \\
& & \mbox{} j_0\left(r'\sqrt{p^2+pqz/2+q^2/16}\right)
j_0\left(r\sqrt{p^2-pqz/2+q^2/16}
\right)\varphi_i(r)\nonumber 
\end{eqnarray}

Here $v_c=p/m_c$, $v_{\bar d}=p/m_{\bar d}$,
and the factor $f_{BS}(p)$ is the ``minimal relativity''
factor~(\ref{bbfact}). 
The expression for
corresponding matrix elements in the case of the $D^0$ and $\bar D^0$
systems is readily inferred from eqn~(\ref{mu_D0}) by comparison. Here
$\varphi_i(r)$ and $\varphi_f(r')$ denote the initial and final orbital
wavefunctions respectively.

It is instructive to compare these expressions to those of the static
quark model $(v_c,v_{\bar d}\rightarrow 0)$. In the case of the $D^\pm$
mesons that expression is simply

\begin{equation}
{\cal M}_{IA}(D^\pm)=\pm \frac{1}
{12}\int_{0}^{\infty}d^3r\: \varphi_f^*(r)\left(\frac{2}{m_c}-\frac{1}
{m_{\bar d}}\right)
j_0\left(\frac{qr}{2}\right)\varphi_i(r)
\label{mu_stat_Dpm}
\end{equation}

whereas for the $D^0,\bar D^0$ systems it is

\begin{equation}
{\cal M}_{IA}(D^0,\bar D^0)=\pm\frac{1}
{6}\int_{0}^{\infty}d^3r\:\varphi_f^*(r)
\left(\frac{1}{m_c}+
\frac{1}{m_{\bar u}}\right)j_0\left(\frac{qr}{2}\right)\varphi_i(r).
\end{equation}

The expressions~(\ref{mu_full_Dpm}) and~(\ref{mu_stat_Dpm}) apply to 
$D^\pm_s$ mesons with the
substitutions $m_{\bar d}\rightarrow m_{\bar s}$ and $v_{\bar
d}\rightarrow v_{\bar s}$.

The matrix elements of the exchange magnetic moment that is associated
with the confining interaction~(\ref{conf}) is in the case of the $D^\pm$
mesons

\begin{equation}
{\cal M}_2(C)=\mp\frac{1}{12}c\int_{0}^{\infty}d^3
r\: r \varphi_f^*(r)\left(\frac{2}{m_c^2}-\frac{1}{m_{\bar
d}^2}\right)j_0\left(\frac{qr}{2}\right)\varphi_i(r).
\end{equation}

This expression applies to the $D^\pm_s$ mesons as well, with the
substitution $m_{\bar d}\rightarrow m_{\bar s}$. In the case of the
$D_0,\bar D_0$ mesons the corresponding expression is

\begin{equation}
{\cal M}_2(C)=\mp\frac{1}{6}c\int_{0}^{\infty}
d^3r\:r\varphi_f^*(r)\left(\frac{1}{m_c^2}+\frac{1}{m_{\bar
u}^2}\right)j_0\left(\frac{qr}{2}\right)\varphi_i(r).
\end{equation}

The matrix element of the gluon exchange spin-flip
magnetic moment operator is in the case of the $D^\pm$
mesons \mbox{(cf.(\ref{mu_ex_Dpm})):}

\begin{equation}
{\cal M}_2(G)=\pm{1 \over 54}\left({2\over m_c^2}
-{1\over m_c m_{\bar d}}
-{1\over m_{\bar d}^2}\right)\int_0^\infty d^3r  \varphi_f^*(r)
{f_0(r)\over r}j_0\left({qr\over 2}\right)\varphi_i(r).
\end{equation}

In the case of the $D^0,\bar D^0$ mesons this matrix
element is

\begin{equation}
{\cal M}_2(G)=\pm{1 \over 27}\left({1\over m_c^2}+
{1\over m_{\bar u}^2}\right)\int_0^\infty d^3r\,\varphi_f ^*(r)
{f_0(r)\over r}j_0\left({qr\over 2}\right)\varphi_i(r).
\end{equation}

The rule for constructing the corresponding matrix
element expressions for the other heavy-light meson
charge states may be inferred from these expressions.

\subsection{M1 transitions of charm mesons}

At present the only experimental information on the M1 transition 
rates of $D$-meson states are the M1 branching ratios for 
$D^*(2010)^\pm \rightarrow D^\pm \gamma$: $1.1_{-0.7}^{+2.1}\%$ and for 
$D^*(2007)^0\rightarrow D^0\gamma$: $38.1\pm2.9\%$ \cite{Caso}. As the 
upper limit for the total width of the former transition is 0.131 MeV 
and for the latter 2.1 MeV, the upper limits for the M1 decay widths 
are accordingly
1.44 keV for $D^*(2010)^\pm \rightarrow D^\pm \gamma$ and 800 keV for 
$D^*(2007)^0\rightarrow D^0\gamma$. These upper limits suggest that 
the M1 decay widths should be of the order 1 keV.

\begin{table}[h!]
\begin{center}
\begin{tabular}{l||c|c|c|c|}
 & NRIA & RIA & RIA + cf. & RIA + cf. + ex. \\ \hline
$D^{*0}\rightarrow D^{0}$ & 20.9 keV & 6.52 keV & 2.21 keV & 1.25 keV \\
$D^{*'0}\rightarrow D^{0}$ & 805 eV & 51.1 keV & 39.3 keV & 44.5 keV \\
$D^{*'0 }\rightarrow D^{'0} $ & 1.88 keV & 631 eV & 2.48 keV & 2.24 keV \\
$D^{'0}\rightarrow D^{*0}$ & 3.02 keV & 35.7 keV & 31.6 keV & 34.2 keV \\
\end{tabular}
\caption{The M1 decay widths of the $D_{0}$ ($c\bar{u}$) mesons} \label{D0_tab}
\end{center}
\end{table}

\begin{table}[h!]
\begin{center}
\begin{tabular}{l||c|c|c|c|}
 & NRIA & RIA & RIA + cf. & RIA + cf. + ex. \\ \hline
$D^{*\pm}\rightarrow D^{\pm}$ & 574 eV & 10.8 eV & 2.09 keV & 1.10 keV \\
$D^{*'\pm}\rightarrow 
D^{\pm}$ & 23.9 eV & 2.77 keV & 12.5 keV & 15.7 keV \\
$D^{*'\pm}\rightarrow D^{'\pm}$ & 52.6 eV & 2.84 eV & 748 eV & 607 eV \\
$D^{'\pm}\rightarrow 
D^{*\pm}$ & 85.3 eV & 1.63 keV & 10.2 keV & 11.9 keV \\
\end{tabular}
\caption{The M1 decay widths of the $D^{\pm}$ ($c\bar{d}$) mesons} 
\label{Dpm_tab}
\end{center}
\end{table}

This is indeed what we find by explicit calculation here. The calculated 
widths for the transitions $D^*(2010)^\pm \rightarrow D^\pm \gamma$
and $D^*(2007)^0\rightarrow D^0\gamma$, as well as for
the transitions
between their respective excited states are listed in 
Tables~\ref{Dpm_tab} and~\ref{D0_tab} 
respectively. The first column gives the nonrelativistic impulse 
approximation results (NRIA), which are obtained with static quark 
current operators. 
The experience from the corresponding decays in charmonium suggests
that these are not realistic. The results obtained
in the relativistic impulse approximation, with the full Dirac magnetic
moment operators are listed in the second column (RIA). These values 
should also, on the basis of the experience with heavy quarkonia, be 
unrealistic. Finally the column RIA + c.f. give the results that are 
obtained when the exchange current contribution that is associated with
the scalar confining interaction is taken into account. The last column
(RIA + c.f. + ex.) also takes into account the contribution of the gluon
exchange current. The contribution of the latter is most significant
for the M1 transitions between the ground states.

In Tables~\ref{Dpm_tab} and~\ref{D0_tab} the M1 transition between the
excited $D^*$ and $D$ 
mesons and the ground states are given as well. In the case of the 
forbidden transitions between the orbitally excites states the two-body
current contributions are dominant. The large difference between
the calculated M1 transition rates of the charged and neutral $D$-mesons
is due to the very different quark mass dependence of the associated 
spin-flip magnetic moment operators. In the derivation of these
results we determined the value of the photon momentum
$q$ from the differences between the empirical values for the
$D$ meson states as given in ref.\cite{Caso}. The difference
is only notable in the case of the excited $D'$ and $D^{'*}$
mesons, the latter of which is
underpredicted by about 35 MeV.

We have also calculated the M1 transition rates for $J/\psi \rightarrow
\eta_c\gamma$ and $\psi'\rightarrow \eta_c\gamma$ with the present
framework and parameters. In this case the gluon exchange current does not
contribute at all because of the equal masses of the charm quarks and 
antiquarks. The empirical width for the $J/\psi\rightarrow \eta_c\gamma$
decay is known to be $1.14 \pm 0.39$ keV. For this we find the value 
0.75 keV when the two-body contribution is included, while the relativistic
impulse approximation value is 1.47 keV. The static quark model prediction
is 2.51 keV, which amounts to an overprediction of about a factor 2.
The empirical width of the forbidden M1 transition $\psi'\rightarrow
\eta_c\gamma$ is $0.78 \pm 0.24$ keV. For this we obtain the value 1.49 
keV when the two-body current is taken into account. This is much closer
to the empirical value than the value 0.17 keV given by the static quark
model or the value 8.48 keV that is obtained in the relativistic impulse
approximation.

The M1 calculated transitions between the $D_s^*$ and $D_s$ mesons are 
listed in Table~\ref{Ds*_tab}. These transition rates should {\it a
priori} be 
expected to be qualitatively similar to those of the $D^{*\pm}$ and $D^\pm$
mesons. This is borne out by the calculation, although the numerical
values show marked differences. This is another illustration of the 
sensitivity of the calculated M1 transition rates to the parameters of the 
model and in this case to the difference between the constituent masses
of the $s$ and $d$ quarks. 

\begin{table}
\begin{center}
\begin{tabular}{l||c|c|c|c|}
 & NRIA & RIA & RIA + cf. & RIA + cf. + ex. \\ \hline
$D_{s}^{*}\rightarrow D_{s}$ & 178 eV & 7.10 meV & 736 eV & 337 eV \\
$D_{s}^{*'}\rightarrow D_{s}$ & 7.88 eV & 1.22 keV & 3.26 keV & 4.47 keV \\
$D_{s}^{*'}\rightarrow D_{s}'$ & 17.1 eV & 246 meV & 258 eV & 200 eV \\
$D_{s}'\rightarrow D_{s}^{*}$ & 22.4 eV & 654 eV & 2.53 keV & 3.13 keV \\
\end{tabular}
\caption{The M1 decay widths of the $D_{s}$ ($c\bar{s}$) mesons}
\label{Ds*_tab}
\end{center}
\end{table}

\subsection{M1 transitions of bottom mesons}

The calculated M1 transition widths of the bottom mesons are
listed in Tables~\ref{B0_tab} and~\ref{Bpm_tab}. The experimental decay
widths 
for these transitions are yet to be measured. The large
masses of the bottom mesons make these transition widths
much smaller than those of the corresponding transitions
of charm mesons.

The overall features of these results are however similar
to those of the charm mesons. The nonrelativistic
impulse approximation leads to considerable overestimates.
In the case of the forbidden transitions the two-body
exchange currents are again dominant.

The calculated M1 transition rates for the strange
bottom mesons are listed in Table~\ref{Bs*_tab}.

\begin{table}[h!]
\begin{center}
\begin{tabular}{l||c|c|c|c|}
 & NRIA & RIA & RIA + cf. & RIA + cf. + ex. \\ \hline
$B^{*0}\rightarrow B^{0}$ & 152 eV & 51.9 eV & 25.1 eV & 9.55 eV \\
$B^{*'0}\rightarrow B^{0}$ & 149 meV & 5.39 keV & 9.73 keV & 12.2 keV \\
$B^{*'0}\rightarrow B^{'0}$ & 27.7 eV & 9.59 eV & 48.3 eV & 40.6 eV \\
$B^{0'}\rightarrow B^{*0}$ & 34.6 eV & 4.31 keV & 7.95 keV & 9.72 keV \\
\end{tabular}
\caption{The M1 decay widths of the $B_{0}$ ($b\bar{d}$) mesons}
\label{B0_tab}
\end{center}
\end{table}

\begin{table}[h!]
\begin{center}
\begin{tabular}{l||c|c|c|c|}
 & NRIA & RIA & RIA + cf. & RIA + cf. + ex. \\ \hline
$B^{*\pm}\rightarrow B^{\pm}$ & 462 eV & 132 eV & 160 eV & 67.4 eV \\
$B^{*'\pm}\rightarrow 
B^{\pm}$ & 455 meV & 20.1 keV & 39.2 keV & 50.9 keV \\
$B^{*'\pm}\rightarrow B^{'\pm}$ & 84.3 eV & 24.6 eV & 221 eV & 183 eV \\
$B^{'\pm}\rightarrow B^{*\pm}$ & 105 eV & 15.8 keV & 32.4 keV & 40.7 keV \\
\end{tabular}
\caption{The M1 decay widths of the $B^{\pm}$ ($b\bar{u}$) mesons}
\label{Bpm_tab}
\end{center}
\end{table}

\begin{table}[h!]
\begin{center}
\begin{tabular}{l||c|c|c|c|}
 & NRIA & RIA & RIA + cf. & RIA + cf. + ex. \\ \hline
$B_{s}^{*}\rightarrow B_{s}$ & 116 eV & 46.4 eV & 945 meV & 148 meV \\
$B_{s}^{*'}\rightarrow B_{s}$ & 2.39 eV & 3.87 keV & 2.29 keV & 3.22 keV \\
$B_{s}^{*'}\rightarrow B_{s}'$ & 20.9 eV & 8.26 eV & 12.4 eV & 9.76 eV \\
$B_{s}'\rightarrow B_{s}^{*}$ & 10.6 eV & 3.12 keV & 1.84 keV & 2.50 keV \\
\end{tabular}
\caption{The M1 decay widths of the $B_{s}$ ($b\bar{s}$) mesons}
\label{Bs*_tab}
\end{center}
\end{table}

\section{Discussion} \label{sec_disc}

\vspace{1cm}

The main result of the present study is that it is
possible to achieve a satisfactory description of the
heavy-light meson spectra, as well as credible
predictions for their radiative spin-flip transitions
with the framework of the Blankenbecler-Sugar equation
with a scalar confining interaction. This conclusion
differs from that in ref.\cite{Ebert}. The calculated 
results are however very sensitive to the parameters
in the model, and to the relativistic damping of the
hyperfine interaction. This suggests that that model
for the hyperfine interaction (relativistic gluon
exchange) may not provide a sufficiently complete
description of the hyperfine interaction. It may
be conjectured that screening the gluon exchange
interaction at shorter distances and compensating
with the instanton induced interaction may lead to
less parameter sensitive results.

Both the Blankenbecler-Sugar and Gross quasipotential
reductions of the Bethe-Salpeter equation lead to
two-body exchange current operators, which connect
to negative energy intermediate states. If the confining
interaction couples as a scalar to the quarks, 
the accompanying two-quark magnetic moment is a pure spin-flip
operator for mesons with equal quark and antiquark masses.
The contribution of this operator is essential for obtaining realistic
M1 transition rates for heavy quarkonia \cite{Timo}. In the
equal mass case the analog two-quark operator that arises
with a vector confining interaction has no spin-flip
component \cite{Tsushima}.

For mesons with unequal quark and antiquark masses the
gluon exchange current operator has a spin-flip
component. It therefore also contributes to the
M1 decay rates of the heavy-light mesons. Although this 
contribution is much smaller than that of
the confining interaction, it is significant in the
case of the M1 transitions between the ground
state vector and pseudoscalar heavy-light mesons, because
of the partial cancellation between the 
matrix elements of the single quark current operators
and those of the exchange current associated
with the confining interaction.

Strong conclusions concerning the M1 transition rates
here have to await experimental determination of the
total width of the heavy-flavor vector mesons. As the
relative branching ratios for M1 decays of these 
mesons are known,
the key missing empirical information is their total
widths for strong decay.
  
\vspace{1cm}

\centerline{\bf Acknowledgments}

\vspace{0.5cm}

This work has been supported in part by Academy of Finland
under contract 43982.


\begin{thebibliography}{99}
\bibitem{Bali} G. S. Bali, K. Schilling and A. Wachter, Phys. Rev.
{\bf 56}, 2566 (1997)
\bibitem{Diak} D. Diakonov and V.Petrov, eprint hep-lat/9810037
\bibitem{Metsch} B. Metsch, Nucl. Phys. {\bf A578}, 418 (1994)
\bibitem{Snellman} J. Linde and H. Snellman, Nucl. Phys. {\bf A619},
346 (1997)
\bibitem{Blankenbecler} R. Blankenbecler and R. Sugar, Phys. Rev. {\bf
142}, 1051 (1966) 
\bibitem{Logunov} A.A. Logunov and A.N. Tavkhelidze, Nuovo Cimento
{\bf 29}, 380 (1963) 
\bibitem{Gross} F. Gross, Phys. Rev. {\bf 186}, 1448 (1969) 
\bibitem{Wally} J. Zeng, J. W. Van Orden and W. Roberts,
Phys. Rev. {\bf D52}, 5229 (1995)
\bibitem{Coe} F. Coester and D. O. Riska, Ann. Phys. {\bf 234},
141 (1994) 
\bibitem{Lomon} M. H. Partovi and E. H. Lomon, Phys. Rev. {\bf D2},
1999 (1980)
\bibitem{Ebert} D. Ebert, V. O. Galkin and R. Faustov, 
Phys. Rev. {\bf D57}, 5663 (1998), {\bf D59}, 1019902 (1999)
\bibitem{Jackson} G. E. Brown, A. D. Jackson and T. T. S. Kuo,
Nucl. Phys. {\bf A133}, 481 (1969)
\bibitem{Amund} J. F. Amundson et al., Phys. Lett. {\bf B296}, 415
 (1992) 
\bibitem{Ang} L. Angelos and 
G. P. Lepage, Phys. Rev. {\bf D45}, R3021 (1992)
\bibitem{Bajc} B. Bajc, S. Fajfer and R. J. Oakes,
Phys. Rev. {\bf D51}, 2230 (1995) 
\bibitem{Timo} T.A. L\"ahde, C. Nyf\"alt and D.O. Riska, 
Nucl. Phys. {\bf A645}, 587 (1999) 
\bibitem{Gromes} D. Gromes, Phys. Lett. {\bf B202}, 262 (1988)  
\bibitem{Godfrey} S. Godfrey and N. Isgur, Phys. Rev. {\bf D32}, 
189 (1985)
\bibitem{Nowak} S. Chernyshev, M.A. Nowak and I. Zahed, Phys. Rev.
{\bf D53}, 5176 (1996)
\bibitem{Rho} M. A. Nowak, M. Rho and I. Zahed, {\it Chiral
Nuclear Dynamics}, World Scientific, Singapore (1996)
\bibitem{Negele1} M.-C. Chu et al., Phys. Rev. {\bf D49}, 6039 (1994)
\bibitem{Negele2} J. W. Negele, Nucl. Phys. Proc. Suppl. {\bf 73},
92 (1999)
\bibitem{Ebert2} D. Ebert, R. N. Faustov and V. O. Galkin,
Eur. Phys. Journal {\bf C7}, 539 (1999)
\bibitem{Risk} D. O. Riska, Nucl. Phys. {\bf B56}, 445 (1973)
\bibitem{Dav1} C. T. H. Davies et al., Phys. Lett. {\bf B345},
42 (1996)
\bibitem{Dav2} C. T. H. Davies et al., Phys. Rev. {\bf D56},
2755 (1997)
\bibitem{Matti} A. C. Mattingly and P. M. Stevenson, Phys. Rev.
{\bf D49}, 437 (1994)
\bibitem{Brod} S. J. Brodsky, C.-R. Ji, A. Pang and 
D. G. Robertson, Phys. Rev. {\bf D57}, 245 (1998) 
\bibitem{Quigg} E. J. Eichten and C. Quigg, Phys. Rev. 
{\bf D52}, 1726 (1995)
\bibitem{Caso} C. Caso et al., Eur. Phys. Journal {\bf C3}, 1 (1998)
\bibitem{Bourd} C. Bourdarios, eprint hep-ex/9811014
\bibitem{Tsushima} K. Tsushima, D. O. Riska and P. G. Blunden,
Nucl. Phys. {\bf A559}, 543 (1993)
\end{thebibliography}
\end{document}